\documentclass[doublecol]{epl2} 

\usepackage{amsmath,amssymb}
\usepackage{placeins}
\usepackage{comment}

\title{Effect of constraint relaxation on dynamic critical phenomena in minimum vertex cover problem}
\shorttitle{Constraint relaxation and dynamic critical phenomena in minimum vertex cover} 

\author{A. Dote\inst{1,3} \thanks{E-mail: \email{dote.aki@fujitsu.com}} \and K. Hukushima\inst{1,2}}
\shortauthor{A. Dote \etal}

\institute{                    
  \inst{1} Graduate School of Arts and Sciences,
  The University of Tokyo -
  Komaba, Meguro-ku, Tokyo 153-8902, Japan\\
  \inst{2} Komaba Institute for Science 
  The University of Tokyo - 3-8-1 Komaba, Meguro-ku, Tokyo 153-8902, Japan\\
  \inst{3} Fujitsu Limited - 
  4-1-1 Kamikodanaka, Nakahara-ku, Kawasaki, 211-8588, Japan
}

\abstract{The effects of constraint relaxation on dynamic critical phenomena in the Minimum Vertex Cover (MVC) problem on Erd\H{o}s-R\'enyi random graphs are investigated using Markov chain Monte Carlo simulations. 
Following our previous work that revealed the reduction of the critical temperature by constraint relaxation based on the penalty function method, this study focuses on investigating the critical properties of the relaxation time along its phase boundary.
It is found that the dynamical correlation function of MVC with respect to the problem size and the constraint strength follows a universal scaling function. 
The analysis shows that the relaxation time decreases as the constraints are relaxed. 
This decrease is more pronounced for the critical amplitude than for the critical exponent, and this result is interpreted in terms of the system's microscopic energy barriers due to the constraint relaxation. 
}

\begin{document}

\maketitle

\section{Introduction}

Combinatorial optimisation problems occupy a crucial position in modern society. 
The advancement of information technology has enabled the gathering of extensive data, making these problems essential for decision-making in diverse fields such as logistics, manufacturing, and finance. 
However, traditional computational methods have become unsuitable for scaling demands. 
In this context, quantum \cite{johnson2011quantum} and classical Ising machines \cite{yamaoka201520k, matsubara2020digital, mcmahon2016fully, goto2019combinatorial, mohseni2022ising} have emerged, offering new possibilities for solving combinatorial optimisation problems and are thus attracting considerable attention.

Practical optimisation problems often involve handling diverse types of constraints.
However, solvers including the Ising machines cannot directly handle these constraints. 
The penalty function \cite{avriel1976nonlinear, bazaraa1979cm} is to efficiently manage such complex constraints. 
This method transforms constrained combinatorial optimisation problems into unconstrained problems by imposing penalties for constraint violations. 
Controlling the strength of these penalties is crucial for finding optimal solutions in a shorter time.

Simulated Annealing (SA)~\cite{kirkpatrick1983optimization} and exchange Monte Carlo (EMC) methods~\cite{hukushima1996exchange}, originating from statistical mechanics, are widely used to solve optimisation problems. 
These methods search for optimal solutions by bridging the system from a high-temperature disordered region to a low-temperature ordered region where optimal solutions are more likely to be found. 
In both the SA and EMC methods, the relaxation time at each temperature is considered to affect the efficiency of the algorithm.
In the case of SA, a theoretically guaranteed but inefficient annealing schedule is known, which requires sufficient relaxation of the system at each temperature \cite{geman1984stochastic}. 
In the EMC method, it has been shown that waiting for the system to sufficiently relax in Markov Chain Monte Carlo (MCMC) at each temperature before exchanging temperatures improves the efficiency of the exchange process \cite{bittner2008make}. 
These suggest that if phase transitions, where the relaxation time diverges in the thermodynamic limit, occur in optimisation problems under consideration, then the critical temperature is a significant bottleneck in these optimisation algorithms.
Moreover, statistical mechanics studies have revealed that phase transitions indeed occur in many optimisation problems defined on random graphs\cite{MezardMontanari,HartmannRieger,HartmannWeigt2005}. 

Statistical mechanical analysis of combinatorial optimisation problems is an effective method for deeply understanding typical properties of these problems, such as their structure and the distribution of solutions. 
In our previous study~\cite{dote2024effect}, we employed the penalty function method in statistical mechanical analysis, focusing on the Minimum Vertex Cover (MVC) problem defined on Erd\H{o}s-R\'enyi (ER) random graphs \cite{erdHos1960evolution}, to reveal the effect of constraint relaxation on problem properties, in particular on the decrease of critical temperature.

This study aims to investigate the effect of constraint relaxation on dynamical critical phenomena through numerical experiments. 
Specifically, we measure the dynamical correlation function numerically, employing MCMC simulations for randomly generated MVC problems over various critical temperatures, which are determined by the strength of the constraints. 
These experiments reveal the dependence of the relaxation time on the problem size and the constraint strength.
As a tool for estimating relaxation times, we have redefined the dimensionless dynamical correlation function to accommodate systems defined by bit variables, expanding its previous application to Ising systems with inversion  symmetry~\cite{bhatt1992new}.

\section{Model}

First, we introduce a statistical mechanical formulation for the MVC problem, which includes constraint relaxation using the penalty function method.

Given an undirected graph $G(V, E)$ with $N$ vertices $V$ and edges $E$, the MVC problem seeks the smallest subset of vertices $V_c \subset V$ that covers $G$. 
A 'vertex cover' refers to a state where at least one vertex $i \in \{1,2,\dots,N\}$ of each edge with $c_{ij}=1$ is included in $V_c$.
Here, $c_{ij} \in \{0,1\}$ represents the elements of the adjacency matrix of the graph $G$.
Denoting $x_i=1$ for $i \in V_c$ and $x_i=0$ for $i \notin V_c$, this problem is formulated as a constrained combinatorial optimisation problem with binary variables $\bm{x} \in \{0,1\}^N$ as follows:
\begin{equation}
    \begin{aligned}
        &\text{minimize: } M(\bm{x}) = \sum_{i=1}^N x_i ,\\
        &\text{subject to: } V(\bm{x}; G) = \sum_{(ij)}c_{ij}(1-x_i)(1-x_j) = 0, 
    \end{aligned}
    \label{eq:problem}
\end{equation}
where $M(\bm{x})$ is the objective function to be minimized, and $V(\bm{x}; G)$ is the penalty function, which takes a positive value when state $\bm{x}$ does not satisfy the constraints and $0$ when it does.
These states are called feasible and infeasible solutions, respectively.

In the penalty function method, an objective function $E(\bm{x};\gamma, G)$ is defined by adding the penalty function to the objective function as follows:
\begin{equation}
    E(\bm{x}; \gamma,G) = M(\bm{x}) + \gamma V(\bm{x}; G), 
    \label{eq:energywithpenalty}
\end{equation}
where $\gamma$ is a positive parameter referred to as the penalty coefficient, which controls the strength of the constraints.  
If the value of $\gamma$ is sufficiently large, the solution $\bm{x}$ that minimizes $E(\bm{x}; \gamma, G)$ satisfies the constraints and becomes the optimal solution to the problem defined in eq.~(\ref{eq:problem}). 
However, if the value of $\gamma$ is too large, the search for a solution remains in the vicinity of the local feasible region, and a globally optimal solution may not be obtained.
It is empirically known that selecting the value of $\gamma$ as small as possible allows efficient search~\cite{smith1997penalty}.
For MVC, $\gamma>1$ is the sufficient condition for obtaining a feasible solution \cite{zhou2003vertex,lucas2014ising,dote2024effect}, although determining its optimal value in advance for a given problem is generally challenging.

We introduce the canonical distribution with eq.~(\ref{eq:energywithpenalty}) as the energy function and the parameter $\beta$ as the 'inverse temperature'. 
The partition function is then expressed as
\begin{equation}
    Z(\beta,\gamma; G) = \sum_{\bm{x}} e^{-\beta (M(\bm{x}) + \gamma V(\bm{x}; G))},
    \label{eq:partitionfunction}
\end{equation}
where the sum is taken for all $2^N$ states $\bm{x} \in \{0,1\}^N$, including feasible and infeasible solutions. 
Note that previous studies \cite{weigt2001minimal,zhang2009stability} restricted to feasible solutions are included as the limit of $\gamma\to\infty$.

As in previous statistical mechanics studies of optimisation problems, this study focuses not on individual problem instances $G$ but rather on the typical properties in a typical ensemble of the instances. 
This approach allows us to utilize the statistical mechanics methods for random systems.  
For the specific typical ensemble, we used the ER random graph \cite{erdHos1960evolution} with a mean degree $c$ as the distribution that the graph $G$ follows.
The elements of the adjacency matrix of $G$ are determined according to the following probability distribution:
\begin{equation}
    P(\{c_{ij}\}) = \prod_{i<j}\left(\frac{c}{N}\delta(c_{ij},1)+\left(1-\frac{c}{N}\right)\delta(c_{ij},0)\right).
\label{eq:ER}
\end{equation}
When $N$ is sufficiently large, the degree distribution follows a Poisson distribution $P(k)=e^{-c}c^k/k!$.
The graph-averaged typical free energy under this distribution has been calculated to discuss the effect of the penalty coefficient $\gamma$ on the system\cite{dote2024effect}.

\section{Phase diagram of MVC}

In studying dynamic critical phenomena, it is necessary to know the phase diagram in MVC. 
The dependence of the critical temperature $T_\mathrm{c}$ on the penalty coefficient $\gamma$ of MVC has been determined in the previous work~\cite{dote2024effect}. 
Similar to the analyses restricted to feasible solutions~\cite{zhang2009stability}, $T_\mathrm{c}$ is estimated by the divergence of the spin-glass susceptibility $\chi_{\rm SG}$, which served as an indicator of replica-symmetry breaking (RSB). 
The spin-glass susceptibility is defined by
\begin{equation}
    \chi_{\rm SG} = \frac{1}{N}\sum_{i\ne j}\left[\langle x_ix_j\rangle_c^2\right]_G ,
    \label{eq:SG_susceptibility}
\end{equation}
where $\langle x_ix_j \rangle_c = \langle x_ix_j \rangle - \langle x_i \rangle\langle x_i \rangle$ is the connected correlation with respect to the thermal average, and $[\dots]_G$ represents the average over $G$. 
Numerical analysis based on the cavity method to evaluate the divergence of $\chi_{\rm SG}$ confirmed that reducing $\gamma$, i.e., relaxing the constraints, leads to a decrease in $T_\mathrm{c}$~\cite{dote2024effect}. 

\begin{figure}[t]
    \centering
    \includegraphics[width=0.75\linewidth]{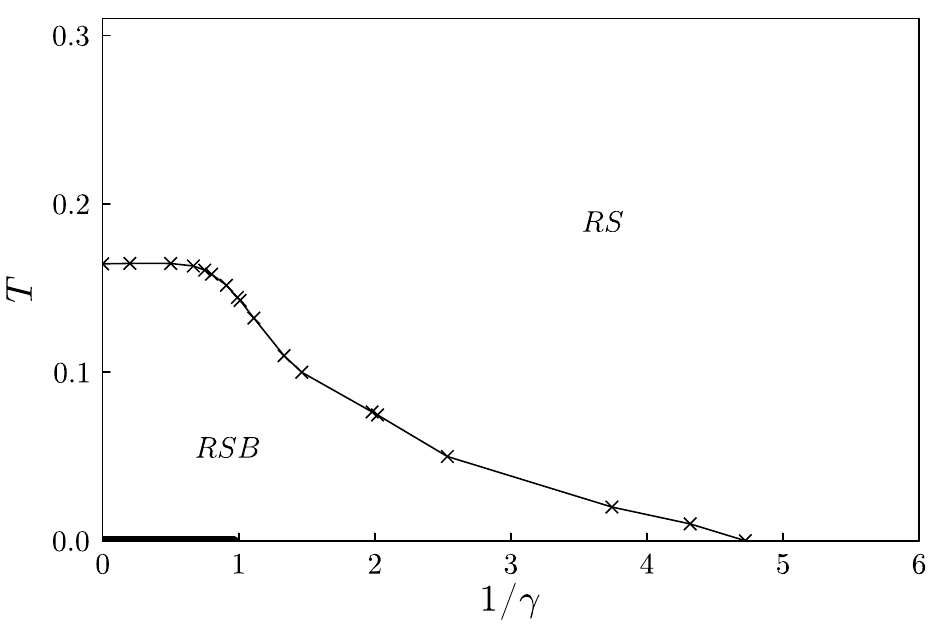}
    \caption{\label{fig:stability_gamma_inv_T_c5} 
    Phase diagram of MVC on the ER random graph with a mean degree $c=5.0$. The critical temperature $T_{\mathrm{c}}$ obtained by the cavity method is denoted by the crosses. 
    This boundary separates the replica symmetry breaking (RSB) phase near the origin side, $T=0$ and $\gamma\to\infty$, from the replica symmetry (RS) phase on the opposite side. The uncertainty in the evaluation of $T_{\mathrm{c}}$ is of the same order as the line thickness and therefore omitted. Note that the markers are connected by straight lines and do not represent exact boundaries. The thick line indicates the region with feasible ground states at $T=0$ and $\gamma>1$. }
\end{figure}

For the ER random graph with a mean degree of $c=5.0$, the dependence of the critical temperature $T_\mathrm{c}$ on the inverse of the penalty coefficient $1/\gamma$ is shown in fig.~\ref{fig:stability_gamma_inv_T_c5}.
The bold line on the horizontal axis represents the range where the system adopts the minimum cover state in the feasible solutions at $T=0$, which is located at $1/\gamma < 1$.
In this figure, the region on the origin side from the phase boundary represents the RSB region, while the outer side represents the RS region. 
The critical temperature $T_\mathrm{c}$ monotonically decreases with $1/\gamma$ and reaches zero at a finite value, indicating the existence of a critical penalty coefficient $\gamma>0$.

\section{MCMC and optimisation algorithms}

MCMC is a method for generating a sample sequence that converges to a target distribution. 
The canonical distribution, which is the target distribution, is defined as 
\begin{equation}
    P(\bm{x};\beta,\gamma, G) = \frac{1}{Z(\beta,\gamma; G)} e^{-\beta E(\bm{x}; \gamma, G)} .
    \label{eq:canonical}
\end{equation}
We apply the simplest form of the Metropolis algorithm.  The algorithm begins with an arbitrary initial state $\bm{x}$ and iteratively operates by flipping a single bit to explore neighbouring states.
Each iteration involves randomly choosing a single bit index $k = 1,2,\dots, N$ and proposing a next state $\bm{x}'$ by flipping the $k$-th bit.
Whether to transition to the proposed state $\bm{x}'$ (accept) or remain in the current state $\bm{x}$ (reject) is determined based on the acceptance probability:
\begin{equation}
    A(\bm{x}\to\bm{x}') = \min(1,\mathrm{e}^{-\beta (E(\bm{x}';\gamma,G)-E(\bm{x};\gamma,G))}).
    \label{eq:acceptance}
\end{equation}
When $\gamma=\infty$, proposals to infeasible solutions are always rejected.
We call $N$ trials as a Monte Carlo step (MCS).

In general, at high temperatures, the acceptance probability in eq.~(\ref{eq:acceptance}) is high, allowing for an extensive exploration of the state space, but the probability of finding the optimal value is very low.
Conversely, at low temperatures, it tends to explore a narrow range of the state space and converge to local minima, making it difficult to escape from these local solutions and find a global optimum. 

SA aims to converge to a global optimum without being trapped in local regions by gradually lowering the temperature while performing MCMC \cite{kirkpatrick1983optimization}. 
This method of lowering the temperature, the annealing schedule, is a crucial factor affecting the efficiency and convergence of the algorithm. 
The schedules that theoretically guarantee convergence to the optimal solution have been proposed, which require MCMC methods to reach equilibrium at each temperature \cite{geman1984stochastic,hajek1988cooling}. 

EMC method\cite{hukushima1996exchange} is primarily a sampling method, but it is also used for optimisation \cite{matsubara2020digital}, particularly effective when the state space is divided into multiple subspaces. 
In this method, copies (replicas) of the same system are  
independently run MCMC at different temperatures, and the temperatures are probabilistically exchanged between replicas at appropriate intervals. 
This exchange process allows each replica to explore a wide area of the state space without being trapped in local minima or regions.
A method has been proposed to optimize the circulation of replicas by adjusting the frequency of exchange attempts according to the length of relaxation time at each temperature \cite{bittner2008make}, further to adjusting the temperature intervals (see review \cite{earl2005parallel}).

As indicated in these studies, the efficiency of optimisation in both SA and EMC methods is deeply connected with the relaxation time at each temperature. 
In these methods, the control parameters should include not only the temperature $\beta$ but also the penalty coefficient $\gamma$ \cite{angelini2019monte}. 
Both of these algorithms bridge a disordered region and a highly constrained region at low temperatures.
As shown in fig.~\ref{fig:stability_gamma_inv_T_c5}, the former corresponds to the RS region, while the latter lies deep within the RSB region, separated by a phase boundary.
On this boundary, the relaxation time for equilibration diverges in the thermodynamic limit, which is considered a major bottleneck in optimisation. 
The aim of this study is to evaluate how the dynamic critical phenomena along this boundary line are influenced by $\gamma$.

\section{Dimensionless dynamical correlation function}

According to the dynamical scaling theory, for sufficiently large time $t$ near the critical temperature $T_{\mathrm{c}}$, the autocorrelation function $C(t)$ of the quantity associated with the order parameter is expected to behave as $C(t)\sim t^{-x} C_0(t/\tau)$ \cite{hohenberg1977theory}, where $x$ is a dynamical critical exponent and $C_0(t/\tau)$ is a scaling function in characteristic time $\tau$. 
In the initial relaxation regime where $t\ll\tau$, the critical power-law dynamics $t^{-x}$ dominates $C(t)$. 
Since $\tau$ tends to diverge at the critical temperature, a large part of the observation occurs during the initial relaxation regime. 
Estimating $\tau$ is challenging because it requires long-time MCMC simulations to evaluate the asymptotic behaviour of $C(t)$, leading to substantial errors and complicating accurate estimation.

To address this, ref.~\cite{bhatt1992new} proposes a dimensionless dynamical correlation function that eliminates the power term by taking the ratio of moments.
Based on their idea, we propose a dimensionless dynamic correlation function applicable to systems without inversion symmetry, including those with the bit variable.
First, the two-time correlation function for a given graph $G$ is defined as
\begin{equation}
    Q(t,t_0) = \frac{1}{N}(\bm{x}(t_0)-\langle\bm{x}\rangle) \cdot (\bm{x}(t_0+t)-\langle\bm{x}\rangle) ,
\end{equation}
where $\langle \cdots \rangle$ represents the thermal average in the system defined by graph $G$. 
The normalized autocorrelation function is expressed as
\begin{equation}
    \Gamma(t;G) = \frac{\langle Q(t,t_0) \rangle}{\langle Q(0,t_0) \rangle}
    =\frac{\langle \bm{x}(t_0)\cdot\bm{x}(t_0+t)\rangle - |\langle \bm{x}\rangle|^2}{\langle |\bm{x}|^2\rangle - |\langle \bm{x}\rangle|^2}, 
\end{equation}
where the equilibration is assumed to have been reached at time $t_0$. 
This function satisfies two normalization conditions: $\Gamma(0;G)=1$ and $\lim_{t\to\infty}\Gamma(t;G)=0$, and is expected to behave as $t^{-x} \Gamma_0(t/\tau)$ for sufficiently large $t$. 
Then, we redefine the dimensionless dynamical correlation function as
\begin{equation}
    R(t) =\frac{\Gamma(t)}{\sqrt{\Gamma_2(t)}}, 
    \label{eq:dimless_autocor}
\end{equation}
where $\Gamma(t) = [\Gamma(t;G)]_G$ and 
\begin{equation}
    \Gamma_2(t) = [\Gamma_2(t;G)]_G = \left[\frac{\langle Q^2(t,t_0) \rangle}{\langle Q^2(0,t_0) \rangle}\right]_G .
\end{equation}
While there is some arbitrariness in how the graph average is taken, the chosen definition ensures that at least the two conditions for $\Gamma(t;G)$ are satisfied for each $G$. 
It can be seen that as $t$ approaches infinity, the limit of $\Gamma_2(t)$ is a positive constant. 
This is because $\langle Q^2 \rangle$ tends to approach $\sum_{i,j}\langle x_i x_j\rangle_c^2 /N^2$ in the limit.
Equation~(\ref{eq:dimless_autocor}) is expected to follow the scaling of $R(t) \sim \tilde{R}(t/\tau)$.
In other words, by applying the scaling with an appropriate $\tau$ depending on $N$, $T$, and $\gamma$, the correlation function $R(t)$ can be characterized by a common universal function $\tilde{R}(t/\tau(N, T,\gamma))$. 
From the naive point of view of the scaling ansatz, this universal scaling function is expected to hold regardless of $N$ and $\gamma$ at $T \sim T_{\mathrm{c}}(\gamma)$, although this ansatz should be verified by numerical simulations.   

\begin{figure}[ht]
    \centering
    \includegraphics[width=0.8\linewidth]{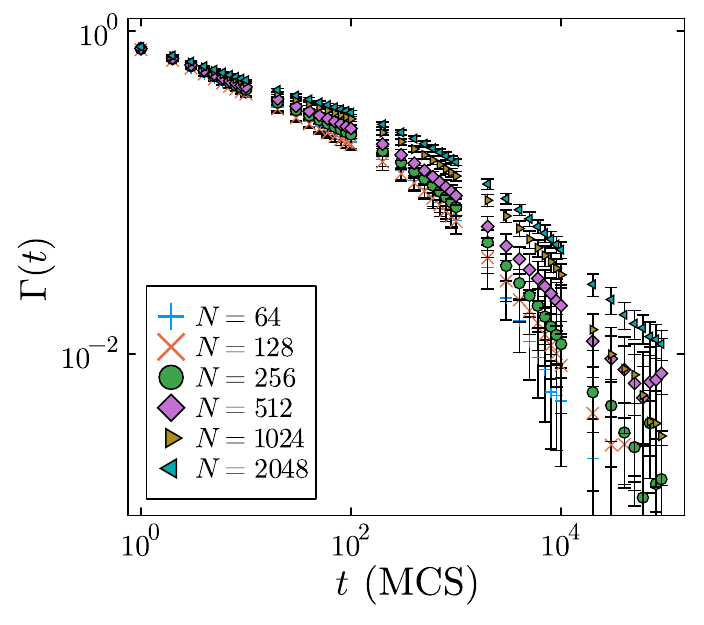}
    \caption{\label{fig:autocor} MCS dependence of the normalized autocorrelation function $\Gamma(t)$ of MVC on ER random graphs with a mean degree $c=5$ and $\gamma=1.1$ at $T_\mathrm{c}=0.152$ for various system sizes $N$. For each $N$, the observations were averaged over $400$--$100$ graphs. The error bar represents the standard error obtained by the bootstrap method.}
\end{figure}

\begin{figure}[ht]
\centering
    \includegraphics[width=0.8\linewidth]{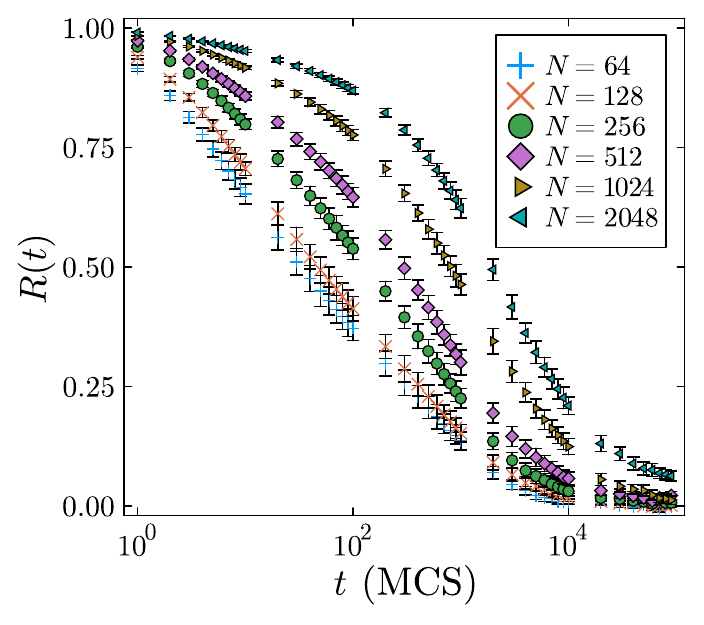}
    \caption{\label{fig:dimless_autocor_N} MCS dependence of the dimensionless dynamical correlation function $R(t)$ of MVC on ER random graphs with a mean degree $c=5.0$ and $\gamma=1.1$ at $T_\mathrm{c}=0.152$ for vaious system sizes $N$. 
    The remaining conditions are the same as in fig. \ref{fig:autocor}.}
\vspace{1mm}
    \centering
    \includegraphics[width=0.8\linewidth]{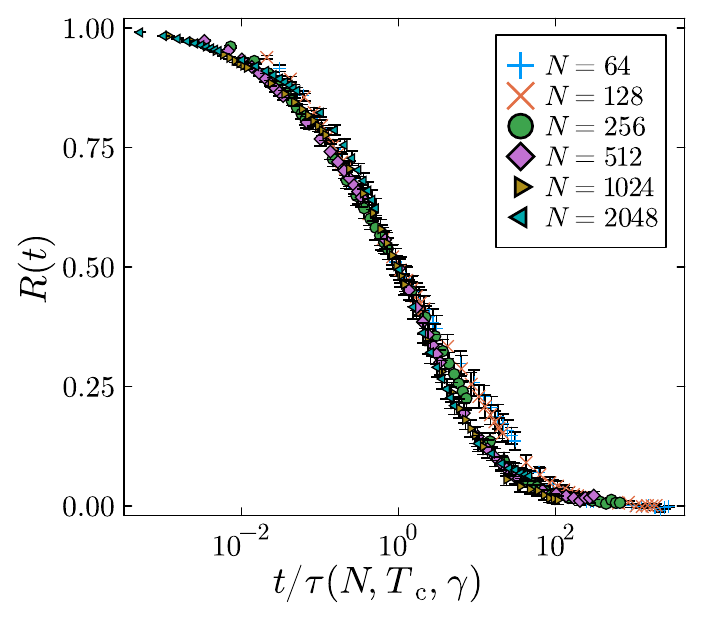}
    \caption{\label{fig:dimless_autocorr_N_rescaled} Scaling plot of the dimensionless dynamical correlation function of MVC on ER random graphs with a mean degree $c=5.0$ and $\gamma=1.1$ at $T_\mathrm{c}=0.152$ for various system sizes $N$. The remaining conditions are the same as in fig. \ref{fig:autocor}.}
\end{figure}

\section{Numerical results}
Here, we focus on the dynamic behaviour at the critical temperature $T_\mathrm{c}$, a major bottleneck of the optimisation algorithms.
Using the dimensionless dynamical correlation function~(\ref{eq:dimless_autocor}), we numerically investigate the penalty coefficient $\gamma$ dependence of the dynamic critical phenomena. 
In the numerical experiments, we first determine $T_\mathrm{c}$ obtained by the cavity method for any given $\gamma$ and set this as the lowest temperature in the EMC simulation, then calculate the expected value $\langle\bm{x}\rangle$ at $T_\mathrm{c}$ by simulations after the system reaches equilibrium. 
Then, using the final state $\bm{x}$ obtained from the simulations as the initial state, we perform the MCMC simulations at $T_\mathrm{c}$ to determine $\Gamma(t;G)$ and $\Gamma_2(t;G)$. 
This procedure of numerical experiments is conducted for randomly generated $G$, and their averages are taken to determine $R(t)$. 
For each size ranging from $N=64$ to $1024$, we averaged a minimum of 100 graphs, increasing up to 400 for $N\le128$ where variations are larger.

As an example, we discuss the results for ER random graphs with $c=5.0$ and $\gamma=1.1$, where $T_\mathrm{c}=0.152$. 
Figure~\ref{fig:autocor} shows the MCS dependence of the normalized autocorrelation function $\Gamma(t)$ for various sizes.
As previously noted, the long-time behaviour of $\Gamma(t)$ needed to estimate $\tau$ has small values and large relative errors due to variations between graphs. 
In contrast, as shown in fig.~\ref{fig:dimless_autocor_N}, the dimensionless dynamical correlation function $R(t)$, which takes values between $0$ and $1$, has a smaller data variation. 
Furthermore, since $R(t)$ is a monotonically decreasing function, and the relaxation time $\tau(N, T,\gamma)$ for given $N$, $T$, and $\gamma$ can be determined from the time at an appropriately selected value of $R(t)$. 

Figure~\ref{fig:dimless_autocorr_N_rescaled} presents the scaling plot for $R(t)$ shown in fig.~\ref{fig:dimless_autocor_N}. 
Although some deviations are observed for $N\le128$, $R(t/\tau)$ scales well over a wide range of $t/\tau$, indicating that the scaling ansatz holds. 
We estimate $N$ dependent $\tau$ at the time when $R(t)=0.5$, but the results are essentially the same when $R(t)=0.2$ or $0.8$ is used, except for the common factor of the relaxation time. 

\begin{figure}[ht]
    \centering
    \includegraphics[width=0.8\linewidth]{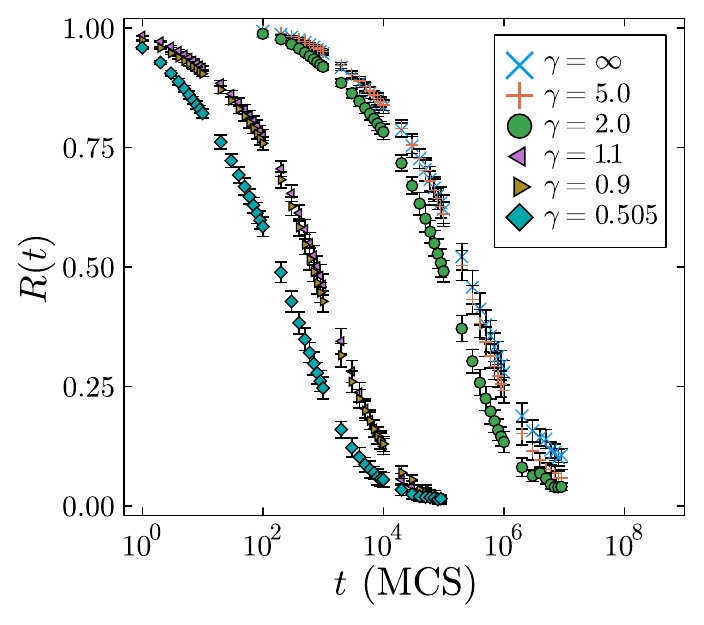}
    \caption{\label{fig:dimless_autocorr_g} MCS dependence of the dimensionless dynamical correlation function $R(t)$ of MVC on ER random graphs with $c=5.0$ and fixed size $N=1024$ for various values of the penalty coefficient $\gamma$. The remaining conditions are the same as in fig.\ref{fig:autocor}.}
    \vspace{5mm}
    \centering
    \includegraphics[width=0.8\linewidth]{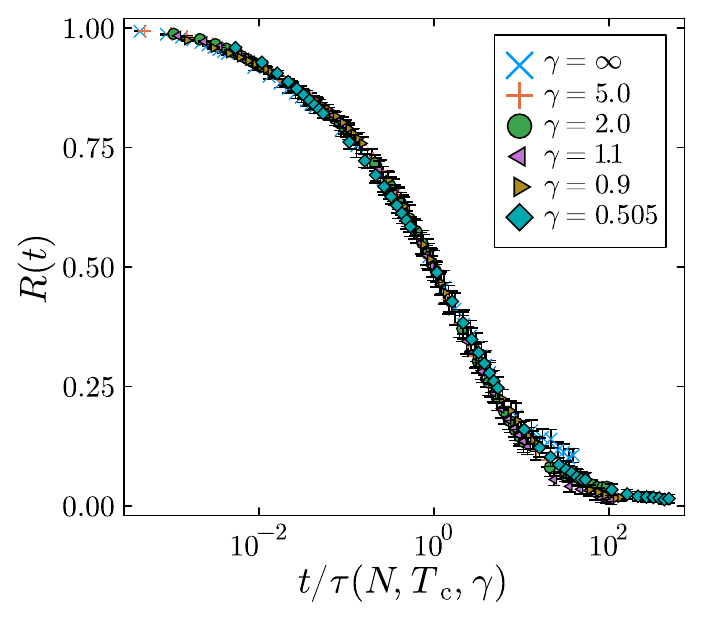}
    \caption{\label{fig:dimless_autocorr_g_rescaled} Scaling plot of the dimensionless dynamical correlation function for various penalty coefficients $\gamma$. The mean degree is $c=5.0$, and the system size is $N=1024$. The remaining conditions are the same as in fig.\ref{fig:autocor}.}
\end{figure}

We also examined whether this scaling holds universally for the penalty coefficient. 
For fixed $N=1024$, $R(t)$ is shown for $\gamma$ ranging from $\infty$ to $0.505$ in fig.~\ref{fig:dimless_autocorr_g}. 
As shown in fig.~\ref{fig:stability_gamma_inv_T_c5}, $T_\mathrm{c}$ depends on $\gamma$, resulting in a different calculated temperature for each $\gamma$.  
It can be observed in fig.~\ref{fig:dimless_autocorr_g} that the relaxation time decreases as $\gamma$ decreases, with a significant decrease between $\gamma=2.0$ and $1.1$.
We will discuss the $N$ and $\gamma$ dependence of the relaxation time in more detail below. 
The scaling plot of $R(t)$, shown in fig.~\ref{fig:dimless_autocorr_g_rescaled}, indicates that it scales well with different $\gamma$.  
This strongly suggests that this scaling function is universal, independent of $\gamma$. 
In addition, the estimated values of $\tau$ remain stable regardless of $N$ and $\gamma$, allowing the determination of relative values for different $\gamma$ without ambiguity. 

\section{Relaxation time and penalty coefficient}
Next, we discuss the asymptotic dynamical behaviour and the effect of $\gamma$ of $\tau(N, T,\gamma)$ obtained from the scaling analysis of the dimensionless dynamical correlation functions.   
In general, the correlation scales typically diverge just above the transition temperature of the second-order phase transition, which is a characteristic inherent to infinite systems. 
In the numerical calculation for finite-size systems, we assume finite-size scaling that explicitly connects the non-divergent behaviour observed in finite-size systems to the divergence behaviour characteristic of infinite systems at $T_c$ \cite{fisher1972scaling}. 
The dynamical finite-size scaling form~\cite{suzuki1977static}, which takes into account the relationship between time and space dimensions, is expressed by 
\begin{equation}
    \tau(N,T_\mathrm{c}(\gamma),\gamma) = a(\gamma) N^{z(\gamma)} ,
    \label{eq:scaling_of_tau}
\end{equation}
where $z(\gamma)$ and $a(\gamma)$ denote the dynamical critical exponent and amplitude, respectively.
This analysis is performed assuming that the value of $T_\mathrm{c}(\gamma)$ is known for given $\gamma$ as shown in fig.~\ref{fig:stability_gamma_inv_T_c5}. 
Note that in a general discussion of dynamical scaling, a length scale is used. 
However, in the context of the ER random graph, where no explicit length scale exists, we use the system size $N$ instead. 
Consequently, it should be recognized that the dynamical exponent differs from the scaling dimension of the time scale with respect to the usual length scale. 
Some arguments suggest that for mean-field models in statistical mechanics without distance structure, the length scaling dimension and the system size are linked by the upper critical dimensions\cite{Botet1982}.
However, for the optimisation problems on random graphs discussed here, the upper critical dimension is not well defined. 

\begin{figure}[t]
    \centering
    \includegraphics[width=0.8\linewidth]{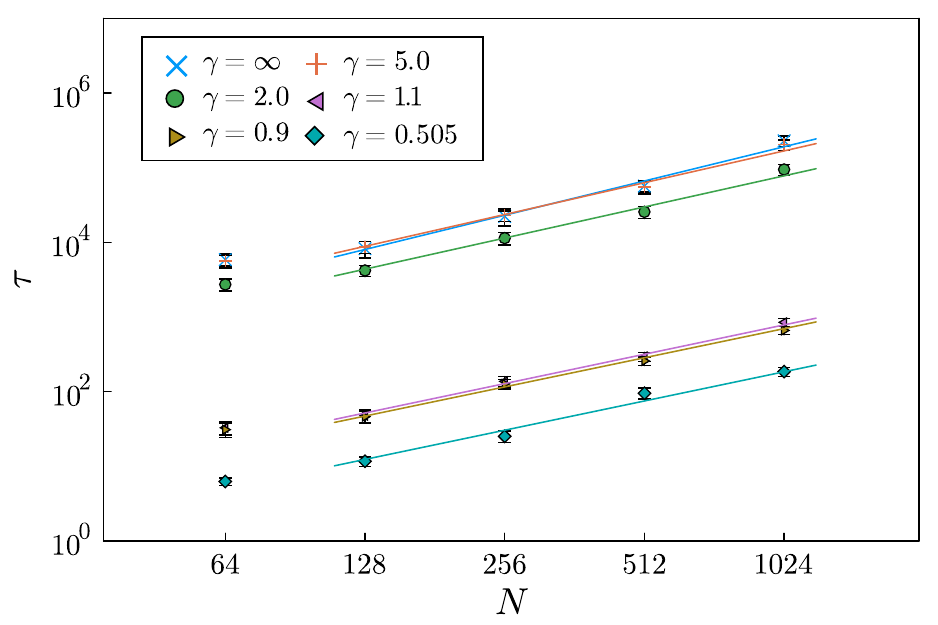}
    \caption{\label{fig:dimless_N_tau}  Dependence of relaxation time $\tau$ on the system size $N$ and penalty coefficient $\gamma$ at the critical temperature $T_\mathrm{c}$, which varies with $\gamma$. The error in $\tau$ was determined using the bootstrap method. The straight lines represent the regression lines fitted to eq.~(\ref{eq:scaling_of_tau}) the least-squares method for each $\gamma$. }
\end{figure}

The system-size $N$ and penalty coefficient $\gamma$ dependence of the relaxation time $\tau$, obtained from the scaling analysis, is shown in fig.~\ref{fig:dimless_N_tau}. 
As mentioned above, since $R(t)$ can be expressed by the universal scaling function for all $\gamma$ examined, this $\gamma$ dependence of the relaxation time is also determined by its relative magnitude. 
It can be seen that for $N\ge128$, $\tau$ follows a power law for large $N$, independent of $\gamma$. 
This means that at the critical temperature where the spin-glass susceptibility diverges, the relaxation times also diverge algebraically with $N$ in the thermodynamic limit.   
It can also be seen that the exponent is almost unchanged with $\gamma$, while the intercept, i.e., the critical amplitude, changes significantly. 

\begin{figure}[h]
    \centering
    \includegraphics[width=0.8\linewidth]{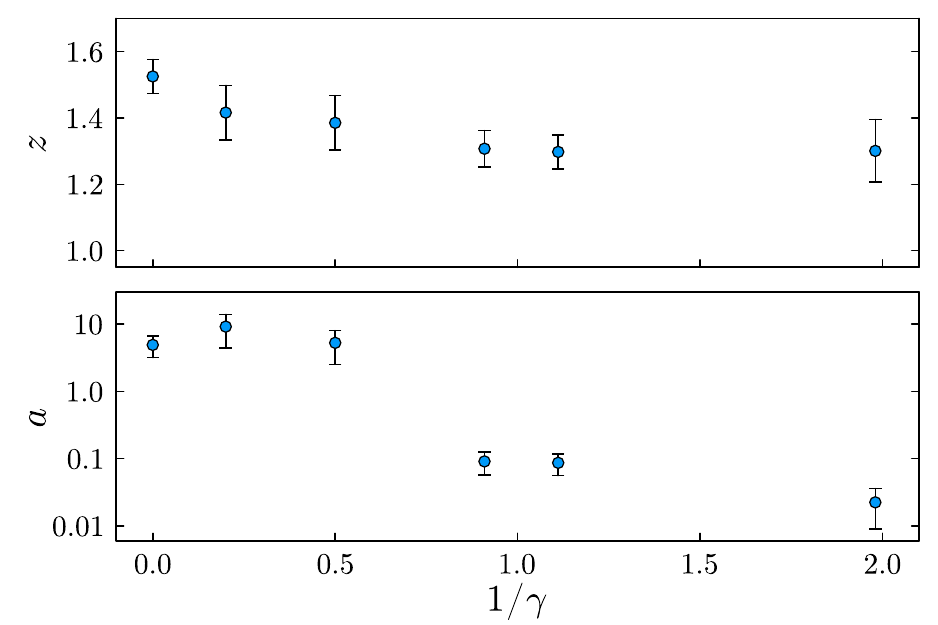}
    \caption{\label{fig:dynamic_critical_amplitude} Dynamic critical exponent $z$ (top) and amplitude $a$ (bottom) as a function of the inverse of the penalty coefficient $1/\gamma$. For each $\gamma$, the values and their associated errors were estimated using the least squares method for the data shown in  fig.~\ref{fig:dimless_N_tau}.}
\end{figure}

The dependence of the evaluated dynamical critical exponent $z$ and amplitude $a$ on $1/\gamma$ is shown in figs.~\ref{fig:dynamic_critical_amplitude} (top) and (bottom), respectively. 
The value of $z$ decreases monotonically from approximately $1.5$ at $1/\gamma=0$, where the system is restricted to the feasible solutions, to about $1.3$ near $1/\gamma=1$, and it remains significantly stable up to $1/\gamma=2.0$. 
In contrast, the value of $a$ does not change significantly from $1/\gamma = 0$ to approximately $0.5$, but it undergoes a significant decrease as $1/\gamma$ approaches $1$, becoming 50 to 100 times smaller.
The critical exponents are generally challenging to evaluate, and it is hard to determine whether the $\gamma$ dependence of $z$ evaluated here is relevant, which is consistent with belonging to the $\gamma$ independent universality class. 
This indicates the need for further research to definitely identify universality classes. 
In contrast to the weak dependence of MVC critical properties on $\gamma$, the amplitude $a$ changes significantly, indicating that the time scale of elementary processes in MCMC strongly depends on $\gamma$.

Intuitively, this change in dynamics can be understood by considering the energy barriers involved in transitions between local minima. 
In MVC, states with the same number of covered vertices differ by at least two bits.
Given a transition between two states that differ by a minimum of two bits, a one-bit flip requires either increasing the cover number by one while maintaining constraints or decreasing the cover number by one and increasing the violation by one. 
The energy increase for these intermediate states, referred to as the energy barriers, is $1$ or $\gamma-1$, respectively.

At relatively low temperatures such as $T_\mathrm{c}$ shown in fig.~\ref{fig:stability_gamma_inv_T_c5}, the minimum energy barrier becomes dominant when states transition according to eq.~(\ref{eq:acceptance}).
For $\gamma\to\infty$, where the transitions are restricted to the feasible solutions, or when $\gamma>2$, the minimum energy barrier is $1$. 
In the range of $1<\gamma<2$, the minimum energy barrier is $\gamma-1<1$.
The lower energy barrier certainly shortens the relaxation time. 
Further relaxation of constraints to $\gamma=1$ results in the energy barrier of $0$, while for $\gamma<1$, infeasible solutions become the minimum. 

This simple argument reasonably explains the critical phenomena, especially the change in the critical amplitude $a$. 
As shown in fig.~\ref{fig:dynamic_critical_amplitude} (bottom), the amplitude $a$ does not change significantly in the range of $\gamma\ge2$, including $\gamma\to\infty$, but decreases significantly for $\gamma<2$, consistent with the change in the minimum energy barrier. 
The change in the critical exponent $z$ is thought to be due to changes in the global structure of the state space and is not affected by local energy barriers.

\section{Conclusion}

In this study, we have investigated the dynamic critical phenomena in MVC on ER random graphs using a 1-bit-flip MCMC method by numerical experiments. 
Applying the penalty function method to clarify the effects of constraint relaxation on critical phenomena presents an interesting challenge from the perspective of solving combinatorial optimisation problems.
Our previous research \cite{dote2024effect} demonstrated that $T_\mathrm{c}$ decreases with the relaxation of constraints. 
In this study, we further have revealed that the relaxation time $\tau$ at $T_\mathrm{c}$ also decreases with the penalty coefficient $\gamma$ decreased. 
In addition, we found that while the critical exponent $z$, which shows the system size $N$ dependence, decreases gradually with the reduction of $\gamma$, the decrease in the $\gamma$-independent critical amplitude $a$ is very significant. 
This significant change in $a$ can be attributed to the reduction of the microscopic energy barriers of MVC caused by the decrease in $\gamma$, and qualitatively, the change in $a$ can be interpreted through this energy barrier change.

In solving combinatorial optimisation problems, it has been empirically suggested that the penalty coefficients should be as small as possible within a range of the coefficients where constraint satisfaction conditions can be obtained. 
Our findings suggest that for MVC, $\gamma\le 2$ is sufficient, supporting the empirical rule.  
However, our evaluation is limited to the phase boundary at the critical temperature, and the structure of the RSB phase below the critical temperature remains unclear. 
Some studies on the cluster structure of the minimum cover states in the low-temperature limit \cite{barthel2004clustering} and on the connectivity of the state space at finite temperatures \cite{sasaki2016numerical} has been conducted, but further detailed analysis regarding solution algorithms is necessary.
In SA and EMC methods, it is possible to vary not only the temperature but also the penalty coefficient, for example, using a path that passes through $\gamma<1$ at relatively high temperatures and reaches $\gamma>1$ at the lowest temperature. The effect of the change in the state space structure when the ground state violates constraints remains unexplored and requires further research. We hope this study contributes to understanding the structure during the search for combinatorial optimisation problems, leading to improvements in existing methods and the development of new ones.

Furthermore, this study introduced a dimensionless dynamical correlation function applicable to systems without inversion symmetry and showed that the same scaling function can be successfully expressed with different $N$ and $\gamma$ for MVC. 
By applying this function to various systems where the dynamical critical phenomena have not been investigated, we expect to obtain novel insights into dynamical critical phenomena in combinatorial optimisation problems.

\acknowledgments
This work was supported by JST Grant Number JPMJPF2221 and JSPS KAKENHI Grant Number 23H01095.

\bibliographystyle{eplbib}
\bibliography{refs_abbrv}

\begin{thebibliography}{10}
\expandafter\ifx\csname url\endcsname\relax\def\url#1{\texttt{#1}}\fi

\bibitem{johnson2011quantum}
\Name{Johnson M.~W., Amin M.~H. \etal} \REVIEW{Nature}{473}{2011}{194}.

\bibitem{yamaoka201520k}
\Name{Yamaoka M., Yoshimura C. \etal} \REVIEW{IEEE J. Solid-State Circuits}{51}{2015}{303}.

\bibitem{matsubara2020digital}
\Name{Matsubara S., Takatsu M. \etal} \Book{Digital annealer for high-speed solving of combinatorial optimization problems and its applications} in proc. of \Book{2020 25th {{Asia S}}. {{Pac}}. {{Des}}. {{Autom}}. {{Conf}}. {{ASP-DAC}}} (IEEE) 2020 pp. 667--672.

\bibitem{mcmahon2016fully}
\Name{McMahon P.~L., Marandi A. \etal} \REVIEW{Science}{354}{2016}{614}.

\bibitem{goto2019combinatorial}
\Name{Goto H., Tatsumura K. \etal} \REVIEW{Sci. Adv.}{5}{2019}{eaav2372}.

\bibitem{mohseni2022ising}
\Name{Mohseni N., McMahon P.~L. \etal} \REVIEW{Nat. Rev. Phys.}{4}{2022}{363}.

\bibitem{avriel1976nonlinear}
\Name{Avriel M.} \Book{Nonlinear programming} in \Book{Mathematical Programming for Operations Researchers and Computer Scientists} (CRC Press) 1976 pp. 271--367.

\bibitem{bazaraa1979cm}
\Name{Bazaraa M.~S., Sherali H.~D. \etal} \Book{Nonlinear Programming: Theory and Algorithms} (John wiley \& sons) 1979.

\bibitem{kirkpatrick1983optimization}
\Name{Kirkpatrick S., Gelatt~Jr C.~D. \etal} \REVIEW{Science}{220}{1983}{671}.

\bibitem{hukushima1996exchange}
\Name{Hukushima K. \and Nemoto K.} \REVIEW{J. Phys. Soc. Jpn.}{65}{1996}{1604}.

\bibitem{geman1984stochastic}
\Name{Geman S. \and Geman D.} \REVIEW{IEEE Trans. Pattern Anal. Mach. Intell.}{}{1984}{721}.

\bibitem{bittner2008make}
\Name{Bittner E., Nu{\ss}baumer A. \etal} \REVIEW{Phys. Rev. Lett.}{101}{2008}{130603}.

\bibitem{MezardMontanari}
\Name{Mezard M. \and Montanari A.} \Book{Information, Physics, and Computation} (Oxford University Press) 2009.

\bibitem{HartmannRieger}
\Name{Hartmann A.~K. \and Rieger H.} \Book{New Optimization Algorithms in Physics} (John Wiley \& Sons, Inc., Hoboken, NJ, USA) 2004.

\bibitem{HartmannWeigt2005}
\Name{Hartmann A.~K. \and Weigt M.} \Book{Phase Transitions in Combinatorial Optimization Problems: Basics, Algorithms and Statistical Mechanics} (John Wiley \& Sons) 2005.

\bibitem{dote2024effect}
\Name{Dote A. \and Hukushima K.} \REVIEW{Phys. Rev. E}{109}{2024}{044304}.

\bibitem{erdHos1960evolution}
\Name{Erd{\H o}s P. \and R{\'e}nyi A.} \REVIEW{Publ. Math. Inst. Hung. Acad. Sci.}{5}{1960}{17}.

\bibitem{bhatt1992new}
\Name{Bhatt R.~N. \and Young A.~P.} \REVIEW{Europhys. Lett.}{20}{1992}{59}.

\bibitem{smith1997penalty}
\Name{Smith A.~E., Coit D.~W. \etal} \Book{Penalty functions} in \Book{Handbook of Evolutionary Computation} (IOP Publishing Ltd., GBR) 1997 p.~C5.

\bibitem{zhou2003vertex}
\Name{Zhou H.} \REVIEW{Eur. Phys. J. B}{32}{2003}{265}.

\bibitem{lucas2014ising}
\Name{Lucas A.} \REVIEW{Front. Phys.}{2}{2014}{5}.

\bibitem{weigt2001minimal}
\Name{Weigt M. \and Hartmann A.~K.} \REVIEW{Phys. Rev. E}{63}{2001}{056127}.

\bibitem{zhang2009stability}
\Name{Zhang P., Zeng Y. \and Zhou H.} \REVIEW{Phys. Rev. E}{80}{2009}{021122}.

\bibitem{hajek1988cooling}
\Name{Hajek B.} \REVIEW{Math. Oper. Res.}{13}{1988}{311}.

\bibitem{earl2005parallel}
\Name{Earl D.~J. \and Deem M.~W.} \REVIEW{Phys. Chem. Chem. Phys.}{7}{2005}{3910}.

\bibitem{angelini2019monte}
\Name{Angelini M.~C. \and {Ricci-Tersenghi} F.} \REVIEW{Phys. Rev. E}{100}{2019}{013302}.

\bibitem{hohenberg1977theory}
\Name{Hohenberg P.~C. \and Halperin B.~I.} \REVIEW{Rev. Mod. Phys.}{49}{1977}{435}.

\bibitem{fisher1972scaling}
\Name{Fisher M.~E. \and Barber M.~N.} \REVIEW{Phys. Rev. Lett.}{28}{1972}{1516}.

\bibitem{suzuki1977static}
\Name{Suzuki M.} \REVIEW{Prog. Theor. Phys.}{58}{1977}{1142}.

\bibitem{Botet1982}
\Name{Botet R., Jullien R. \etal} \REVIEW{Phys. Rev. Lett.}{49}{1982}{478}.

\bibitem{barthel2004clustering}
\Name{Barthel W. \and Hartmann A.~K.} \REVIEW{Phys. Rev. E}{70}{2004}{066120}.

\bibitem{sasaki2016numerical}
\Name{Sasaki M. \and Hukushima K.} \REVIEW{J. Phys. Soc. Jpn.}{85}{2016}{074602}.

\end{thebibliography}

\end{document}